\documentstyle[12pt,aaspp4]{article}

\begin{document}

\title{A Reanalysis of Small Scale Velocity Dispersion in the CfA1
Survey}

\author{Rachel S. Somerville\altaffilmark{1}, Marc
Davis\altaffilmark{2}, and Joel R. Primack\altaffilmark{1}}

\altaffiltext{1}{Physics Department, University
of California, Santa Cruz, CA 95064} 
\altaffiltext{2}{Departments of Astronomy and Physics, University of
California, Berkeley, CA 94720}

\begin{abstract}
The velocity dispersion of galaxies on scales of $r\sim1h^{-1}$ Mpc,
$\sigma_{12}(r)$, may be estimated from the anisotropy of the
galaxy-galaxy correlation function in redshift space. We present a
reanalysis of the CfA1 survey, correct an error in the original
analysis of Davis and Peebles (1983), and find that $\sigma_{12}(r)$
is extremely sensitive to the details of how corrections for infall
into the Virgo cluster are applied. We conclude that a robust value of
$\sigma_{12}$ cannot be obtained from this survey. We also discuss
results from other redshift surveys, including the effect of removing
clusters.
\end{abstract}


\section{Introduction}

Davis and Peebles (1983, hereafter DP83) calculated the velocity
dispersion of galaxies, $\sigma_{12}(r)$, on scales of $r \sim 1 - 5
h^{-1} Mpc$ for the CfA1 redshift survey, a survey containing 1840
redshifts covering 1.83 steradians in the North galactic hemisphere
(Huchra et al.  1983). Their result, $\sigma_{12}(1) = 340\pm40$ km/s
on scales of $1 h^{-1}$ Mpc, became the standard by which N-body
simulations were judged for perhaps ten years, and a primary argument
against the Cold Dark Matter scenario for structure formation with the
assumption that galaxies trace the mass fluctuations in an unbiased
way, which yields much higher velocities on this scale (e.g., Davis et
al. 1985, Gelb
\& Bertschinger 1994). The same calculation was done on the Southern
Sky Redshift Survey (SSRS1, da Costa et al. 1991), with results of
$\sigma_{12}(1)\sim 300$ km/s (Davis 1988), in apparent agreement with
the CfA1 result. It is only recently that there have been attempts to
reproduce the results of DP83 (Mo, Jing, \& Borner 1993, Zurek et
al. 1994), and to perform this analysis on new, larger redshift
surveys (Fisher et al. 1994a, 1994b, Marzke et al. 1995, Guzzo et
al. 1995). It is now apparent that there is a large variation in
$\sigma_{12}$ between different surveys (see Table 1). In addition,
different workers (Mo et al. 1993, Zurek et al. 1994) obtain very
different results (from DP83 and from each other) when they analyze
the CfA1 survey. In this paper we clarify some details of the original
calculation of DP83 which were not spelled out in the original paper,
and present a reanalysis which shows why the results are so
unstable. We also reproduce the earlier results for SSRS1 (Davis
1988), and investigate how removing clusters affects $\sigma_{12}$.

\section{Method}
\label{sec-method}
In this section we briefly describe the method used to extract the pairwise
velocity dispersion $\sigma_{12}$ from the redshift-space correlation
function $\xi(r_{p}, \pi)$. Readers should refer to DP83 and Fisher et
al. (1994a, b) for more details. 
 
The correlation function in redshift space, $\xi(r_{p},
\pi)$, is estimated by counting the number of pairs in a bin in
$r_{p}$ (separation perpendicular to the line of sight) and $\pi$
(separation parallel to the line of sight). It is normalized by
constructing a catalog of Poisson distributed points with the same
selection function and angular limits as the data, and counting pairs
between the data and the Poisson catalog.
\begin{equation}
1+\xi(r_{p}, \pi) = \frac{n_{R}}{n_{D}} \cdot \frac{DD(r_{p},
\pi)}{DR(r_{p}, \pi)}   \label{eq:xi}
\end{equation}
where DD is the number of pairs between data and data, and DR is
number of pairs between the data catalog and a Poisson catalog. The
quantities $n_{R}$ and $n_{D}$ are the minimum variance weighted
densities (see Davis \& Huchra 1982) of the Poisson and data
catalogs, respectively.

Let $F(\bf{w \mid r})$ be the distribution function of velocity
differences $\bf{w}$ for pairs of galaxies with vector separation
$\bf{r}$, and $f(w_{3} \mid r)$ the velocity distribution function
averaged over the directions perpendicular to the line of sight.  The
first moment of $F(\bf{w \mid r})$, $\overline{v_{12}}(r)$, is the
mean streaming velocity relative to the Hubble flow, and from isotropy
it must be a function only of the magnitude of
$\bf{r}$. Correspondingly, the first moment of $f(w_{3} \mid r)$ is
$\langle w_{3} \rangle = y \:
\overline{v_{12}}(r)/r$ where $y$ is the component of $\bf{r}$ along the
line of sight.  

A reasonable form for the distribution function $f(w_3 \mid r)$,
parameterized by its moments, is
\begin{equation}
f(w_3 \mid r) \propto \exp 
\left( \nu \left[\frac{w_3(r)-\langle w_3(r) \rangle}
{\sigma_{12}(r)}\right]^{n}\right)
\end{equation}
It has been found emperically from studying observations and N-body
simulations (Peebles 1976, Fisher et al. 1994b, Zurek et al. 1994,
Marzke et al. 1995) that on small scales an exponential form (n=1)
fits the data better than a Gaussian (n=2) or any higher power of the
argument.  Adopting this form for $f(w_3
\mid r)$, and using $r^{2} = r_{p}^{2} + y^{2}$ and $w_3 = \pi - y$ we
have:
\begin{equation}
1 + \xi(r_{p}, \pi) = \frac{H_{0}}{\sqrt{2}}
\int \frac{dy}{\sigma_{12}(r)} \left[1 +
\xi(r)\right] \exp \left[\frac{-\sqrt{2}}{\sigma_{12}(r)}
 \left| \pi- H_0 y \left[1 - \frac{\overline{v_{12}}(r)}{H_{0}r}
\right] \right| \right]   \label{eq:modelxi}
\end{equation}
An approximation based on self-similar solutions of the BBGKY
hierarachy suggests a form for $\overline{v_{12}}(r)$ (Davis \&
Peebles 1977):
\begin{equation}
\overline{v_{12}}(r) = \frac{FH_{0}r}{[1 + (r/r_{0})^{2}]}  \label{eq:infall}
\end{equation}
where $F$ is an adjustable parameter of the model. The assumption of
stable clustering (that the collapse of the cluster is exactly
balanced by the Hubble flow) leads to F = 1. DP83 investigated
different values of F but the usually quoted result is for F=1. The
velocity dispersion $\sigma_{12}(r)$ is obtained by fitting the model
of equation~(\ref{eq:modelxi}) to $\xi(r_{p},\pi)$ as estimated from
equation~(\ref{eq:xi}).

\section{Reanalysis of CfA1}

In the CfA1 survey, the Virgo cluster dominates the foreground of the
sample. At the time when the analysis of DP83 was done, an accepted
approach to correcting for the infall of our galaxy and other galaxies
into Virgo was to assume a spherical infall model. In this
model the infall velocity of any galaxy was taken to be inversely
proportional to the galaxy's distance from Virgo and the model was
scaled so that the infall of the Local Group had the then-favored
value of 440 km/s (currently favored values are 200-300 km/s).
Therefore in DP83 the redshift of each galaxy was corrected both for
the infall of the local group, and for its own infall into Virgo. Each
redshift was also corrected for galactic rotation using a rotation
velocity of 220 km/s. We will subsequently refer to the correction to
each galaxy's distance due to the infall of the local group towards
Virgo, plus the galactic rotation correction, as the ``dipole''
correction, and the correction due to each galaxy's own infall into
Virgo (plus galactic rotation and the dipole correction) as the
``inverse-Virgo-distance'' or $1/r_{V}$ correction.

There are two stages of the procedure in which these corrections may
be applied. First, there is the distance used in calculating an
absolute magnitude, used to create a semi-volume limited catalog by
eliminatating all galaxies with absolute magnitude $M_B > -18.5 +
5\log h$. Second, there is the velocity used in calculating the
correlation function $\xi(r_p, \pi)$. The choice of which corrections
are to be used in each of these quantities is important, as we shall
see. 

It could be argued that infall corrections should only be
applied to the velocities used as distances for the volume
limiting. The method we use here to extract $\sigma_{12}$ relies on
measuring the distortion of $\xi(r_p, \pi)$ due to peculiar
velocities, so in general one would not want to use corrected
velocities in the pair counting to compute $\xi(r_p, \pi)$.  However,
in the CfA1 sample the shear due to the presence of the Virgo cluster
in the foreground is so large that it could be argued that some
correction is needed even in the velocities used in pair-counting.
   
Because the inverse-Virgo-distance model is singular near Virgo, DP83
applied the infall corrections differently depending on which region
of the survey a galaxy was in. Define the ``inner Virgo core'' as all
galaxies with $\theta_{V} < 6^{\circ}$ and $v < 2500$ km/s, where
$\theta_{V}$ is the angle with respect to the center of the Virgo
cluster ($\alpha_{V} = 12^{h}.48$, $\delta_{V} = 12.67^{\circ}$) and
$v$ is the redshift, the ``outer core'' as galaxies with $6^{\circ} <
\theta_{V} < 14^{\circ}$ and $v < 2500$, and the ``field'' as all the
remaining galaxies. Note that the determination of whether a galaxy is
assigned to the Virgo core depends on which of the velocities
(uncorrected, dipole corrected or inverse-Virgo-distance corrected) is
used for the velocity condition. DP83 used the inverse-Virgo-corrected
velocity.

Table 2 defines various permutations of ``cuts'', or ways of applying
the different corrections in the different regions of the survey. The
entry Cut~1 shows the choices necessary to reproduce the original
results of DP83. This reflects our discovery, in the course of this
reanalysis, that the inner core of Virgo was inadvertantly deleted in
the original analysis due to a typographical error in the computer
program used to obtain the results presented in DP83. The entry Cut~2
is what DP83 intended to do: the mean velocity of the Virgo cluster
($\langle v_{virgo} \rangle$ = 1460 km/s) is assigned to all galaxies
in the inner core for the volume limiting, the dipole corrected
velocities are used in pair counting in the inner core, and in both
the volume limiting and pair counting in the outer core, and the
inverse-Virgo-distance corrected velocities are used in the field.
Note that this value of $\langle v_{virgo} \rangle$ is the measured
mean redshift of Virgo, 1020 km/s, plus the dipole correction for the
infall of the Local Group, 440 km/s.  Cut~3 is the same as Cut~2
except that the inverse-Virgo-distance correction is not used in the
pair counting.  Cut~4 uses the mean Virgo redshift for the volume
limiting in the inner core, but uses only the dipole correction in all
the other regions. This is to enable us to disentangle the effect of
using the mean Virgo velocity in the inner core from the effect of the
inverse-Virgo-distance correction.

We now know that a simple spherical infall model is not a very good
model for the flow field around the Virgo cluster. Even in the
relatively low mass Virgo cluster favored by the Faber-Burstein Great
Attractor flow model, the redshift-distance relation is triple valued
within $20^\circ$ of Virgo (Nolthenius 1993). In order to study the
effects of using of the spherical infall model, Cut~5 and Cut~6
include only the dipole correction in all the regions of the
survey. Cut~5 includes the correction in the pair counting whereas
Cut~6 does not. Finally, we show the result obtained if no corrections
at all are applied to the redshifts at any stage in the procedure. We
also show the result where no corrections are applied but all clusters
with internal velocity dispersion greater than 500 km/s have been
removed using an automated cluster-removing program (see Somerville,
Primack, \& Nolthenius 1996).

Figure~\ref{fig:dp83} shows the results for $\sigma_{12}(r)$ for each
of the cuts. As can be seen the results are very sensitive to the way
in which the cuts are applied. Note that the scale dependence of
$\sigma_{12}$ with $r$ also changes significantly depending on the way
the corrections are applied. In all cases the slope is steeper than
the original result of DP83.  One can understand these results
qualitatively as follows. In Cut~6, the dipole correction assigns
larger distances to all the galaxies, especially those in the Virgo
core. This makes the galaxies seem more luminous and more of them make
it into the volume-limited catalog. In Cut~6, 42 galaxies from the
Virgo core are included in the volume-limited catalog, as compared to
29 when no corrections are used. The galaxies in the Virgo core have
large pairwise velocities, and $\sigma_{12}$ is pair-weighted, so this
increases $\sigma_{12}$ considerably. Cut~5 consists of the same
galaxies as Cut~6, but using the dipole corrected velocities to
calculate $\xi(r_p, \pi)$ reduces the velocity dispersion of the Virgo
cluster and hence the measured $\sigma_{12}$ for the sample. The
differences in $\sigma_{12}$ for Cut~3, Cut~4, and Cut~5 are not
significant - the correlation functions appear very similar, and it is
only because the correlation functions are very noisy at large $\pi$
that different values of $\sigma_{12}$ are obtained. The values are
within the (large) formal errors on the fit. This implies that using
the mean Virgo redshift for the galaxies in the inner core of Virgo
did not have a large effect on the results. In Cut~2, galaxies with
$v>2500$ km/s, the cutoff for the Virgo core, are assigned larger
velocities due to the $1/r_{V}$ correction. This whole band of
galaxies is therefore moved farther away from the Virgo core, reducing
the number of pairs containing Virgo galaxies which fall in the 1
$h^{-1}$ bin, and reducing $\sigma_{12}(1)$. Finally, not
surprisingly, removing the Virgo core altogether reduces $\sigma_{12}$
significantly, as can be seen in Cut~1.

Our analysis of the SSRS1 survey yields $\sigma_{12}(1) \sim 323 \pm
91 $ km/s (SSRS1), which is in agreement with previous results (Davis
1988, Mo et al. 1993). The SSRS1 survey does not contain any rich
foreground clusters, and therefore the original analysis was not
complicated by any infall corrections. Unlike CfA1, when we remove the
clusters from SSRS1 using the cluster-removal routine, the value of
$\sigma_{12}(1)$ does not change significantly (see
Figure~\ref{fig:sigdata}). This suggests that $\sigma_{12}$ for CfA1
is dominated by the Virgo cluster and that the SSRS1 value is more
typical of the field. However, the analysis of existing redshift
surveys does not yet allow a definite conclusion. Removing clusters
appears to reduce sample-to-sample variation in $\sigma_{12}$ (see
Table 3) but work on simulations (Somerville et al. 1996) suggests
that this will also reduce the ability of the statistic to
discriminate between different cosmological models.

\section{Conclusions}

We hope that this paper has removed the confusion regarding the value
of the velocity dispersion in the CfA1 survey. We have shown that the
value of the velocity dispersion is extremely sensitive to the way in
which corrections for infall into the Virgo cluster are applied. This
is because the Virgo cluster is in the foreground of the CfA1 survey
and contains many intrinsically faint galaxies in a thermally hot
region.  Increasing the distance to these galaxies by a small amount
results in inclusion of fewer of these galaxies in the volume limited
sample. Because $\sigma_{12}$ is pair weighted, including or leaving
out even $\sim10$ galaxies from the Virgo cluster can change
$\sigma_{12}$ by $\sim 100-200$ km/s. In addition, including
corrections for cluster infall in the calculation of the correlation
function in redshift space, $\xi(r_{p}, \pi)$, effectively removes
part of the ``finger of god'' and reduces the velocity dispersion of
the cluster.  Once again the pair-weighted nature of the statistic
means that this will result in a significant reduction in the overall
value of $\sigma_{12}$ for the sample.

However, no infall corrections were used in recent calculations of
$\sigma_{12}$ for other redshift surveys and yet a wide range of
values for $\sigma_{12}$ is obtained for different surveys. We have
argued that this is because $\sigma_{12}$ is extremely sensitive to
clusters, and existing redshift surveys do not sample a large enough
volume of space to represent a fair sample of these relatively rare
objects. One approach to solving this problem is to remove the
clusters from the sample before calculating $\sigma_{12}$; however,
this reduces the ability of the statistic to discriminate between
cosmological models. In addition, our work suggests that the results
are likely to be sensitive to the details of how the clusters are
identified and removed. Analysis of larger volume redshift surveys
will be necessary in order to obtain a robust value of $\sigma_{12}$
which is useful in discriminating between cosmological models or for
estimating $\Omega_{0}$. In the meantime, modified velocity statistics
such as the galaxy-weighted velocity dispersion (Miller et al. 1996),
a density dependent version of the pairwise velocity dispersion
(Strauss 1996), or the median velocity of groups (Nolthenius, Klypin
\& Primack 1996) have been designed to be less sensitive to clusters
and may be promising alternatives.

\clearpage

\begin{center}
Acknowledgements
\end{center}

\begin{acknowledgements}
We thank R. Nolthenius for useful comments on the text. RSS
acknowledges support from the NSF in the form of a GAANN fellowship.
MD and JRP are supported by grants from the NSF and NASA.
\end{acknowledgements}

\clearpage

\begin{table*}
\begin{tabular}{cccc}
Survey & $\sigma_{12}(1 h^{-1} Mpc)$ & Reference & Comments\\
\tableline
CfA1 & $340 \pm 40$ km/s & Davis \& Peebles 1983 \\
CfA1 & $276 \pm 17$ km/s & Mo et al. 1993 & with infall correction\\
CfA1 & $433 \pm 24$ km/s & Mo et al. 1993 & no infall correction\\
CfA1 & $\sim 540$ km/s & Zurek et al. 1994 & with infall correction\\
CfA1 & $\sim 580$ km/s & Zurek et al. 1994 & no infall correction\\
CfA1 & $\sim 450-870$ km/s & Somerville et al. 1996 & see text\\
SSRS1 & $\sim 300$ km/s & Davis 1988 & \\
SSRS1 & $242 \pm 28$ km/s & Mo et al. 1993 \\
SSRS1 & $323 \pm 91 \pm 65$ km/s & Somerville et al. 1996 \\
IRAS & $317^{+40}_{-49}$ km/s & Fisher et al. 1994b \\
CfA2 North & $647 \pm 52$ km/s & Marzke et al. 1995 \\
CfA2 South & $367 \pm 38$ km/s & Marzke et al. 1995 \\
SSRS2 & $272 \pm 42$ km/s & Marzke et al. 1995 \\
Perseus-Pisces & $769^{+171}_{-342}$ km/s & Guzzo et al. 1995 \\

\end{tabular}

\caption{Velocity dispersions at 1 $h^{-1}$ Mpc for various redshift surveys.}

\end{table*}

\clearpage

\begin{table*}
\begin{tabular}{|c|c|c|c|c|c|}
\tableline
 & volume limit & pair count & $\sigma_{12}(1)$ (km/s)&$N_{gal}$ &$N_{virgo}$\\
\tableline
cut 1 & & & $346 \pm 98$ & 1235 & 0\\
inner core & deleted & deleted & & &\\
outer core & dipole & dipole & & &\\
field      & $\frac{1}{r_{V}}$ & $\frac{1}{r_{V}}$ & & &\\
\tableline
cut 2 & & & $453 \pm 118$ & 1268 & 33 \\ 
inner core & $\langle v_{virgo} \rangle$ & dipole & & &\\ 
outer core & dipole & dipole & & &\\
field      & $\frac{1}{r_{V}}$ & $\frac{1}{r_{V}}$ & & &\\
\tableline
cut 3 & & & $666 \pm 189$ & 1268 & 33\\ 
inner core & $\langle v_{virgo} \rangle$ & dipole & & &\\ 
outer core & dipole & dipole & & &\\
field      & $\frac{1}{r_{V}}$ & dipole & & &\\
\tableline
cut 4 & & & $737 \pm 229$ & 1195 & 46\\  
inner core & $\langle v_{virgo} \rangle$ & dipole & & &\\ 
outer core & dipole & dipole & & &\\
field      & dipole & dipole & & &\\
\tableline
cut 5 & dipole & dipole & $646 \pm 184$ & 1191 & 42\\  
\tableline
cut 6 & dipole & none & $867 \pm 233$ & 1191 & 42 \\
\tableline
no corrections & none & none & $618 \pm 113$ & 1021 & 29 \\
\tableline
clusters removed & none & none & $406 \pm 85 $ & 972 & 0\\
\tableline
\end{tabular}

\caption{Different ways of applying Virgo infall corrections.
$N_{gal}$ is the number of galaxies included in the volume-limited
catalog. $N_{virgo}$ is the number of galaxies in the Virgo core
included in the volume-limited catalog. See the text for definition of
the labels. }

\end{table*}

\clearpage

\begin{table*}
\begin{tabular}{cccc}
Survey & $\sigma_{12}(1 h^{-1} Mpc)$ & Reference \\
\tableline
CfA1 & $406 \pm 85$ km/s & Somerville et al. 1996 \\
SSRS1 & $321 \pm 90$ km/s & Somerville et al. 1996 \\
IRAS & $317^{+40}_{-49}$ km/s & Fisher et al. 1994b \\ 
CfA2 North & $ 388 \pm 56$ km/s & Marzke et al. 1995 \\
CfA2 South & $ 253 \pm 54$ km/s & Marzke et al. 1995 \\
SSRS2 & $ 275 \pm 64$ km/s & Marzke et al. 1995 \\
Perseus-Pisces & $613^{+73}_{-57} $ km/s & Guzzo et al. 1995 \\

\end{tabular}

\caption{Velocity dispersions at 1 $h^{-1}$ Mpc for various redshift
surveys, with clusters removed. Clusters with internal velocity
dispersion greater than 500 km/s were removed from CfA1 and SSRS1
using an automated computer routine. IRAS galaxies avoid rich clusters
so no cluster removal is necessary. Clusters with Abell richness $R
\ge 1$ were removed from the CfA2/SSRS2 data by hand. In
Perseus-Pisces, the sample was restricted to $RA \le 3^{h}10^{m}$,
thus excluding the Perseus cluster, which is the richest cluster in
the survey. }

\end{table*}

\clearpage

\begin{figure} 
\caption{The velocity dispersion $\sigma_{12}$ for different ways of applying
the corrections for Virgo infall. Cut~1 reproduces the results of
DP83. See Table 2 and the text for a summary of the ``cuts''.  }
\label{fig:dp83}
\end {figure}

\begin{figure} 
\caption{The velocity dispersion
$\sigma_{12}$ for CfA1 and SSRS1, without any corrections for infall
towards Virgo. Filled squares are results for CfA1, open squares are
CfA1 with clusters removed using an automated cluster removing
procedure. Filled triangles are results for SSRS1, open triangles are
SSRS1 with clusters removed.  }
\label{fig:sigdata}
\end{figure}

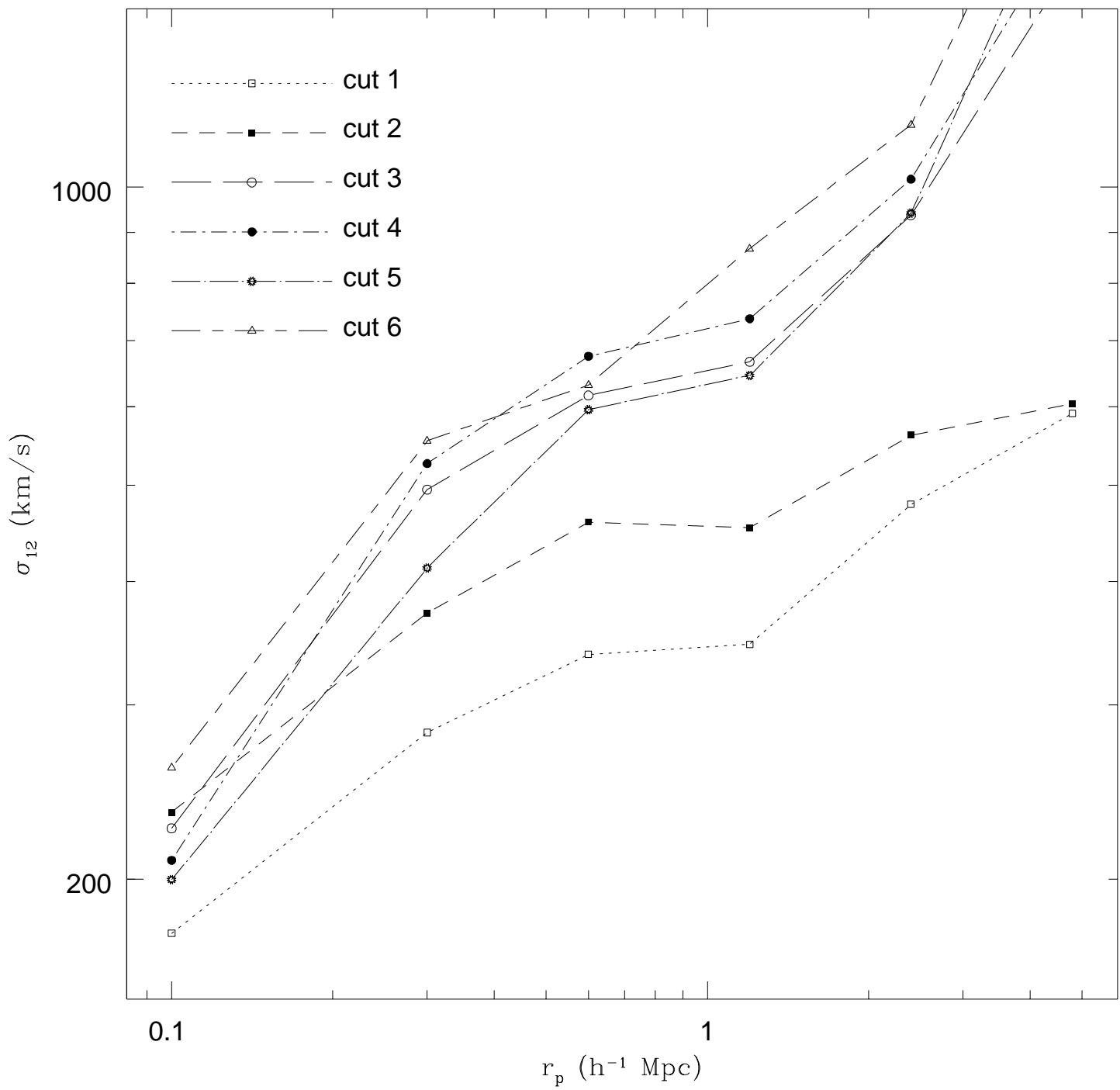
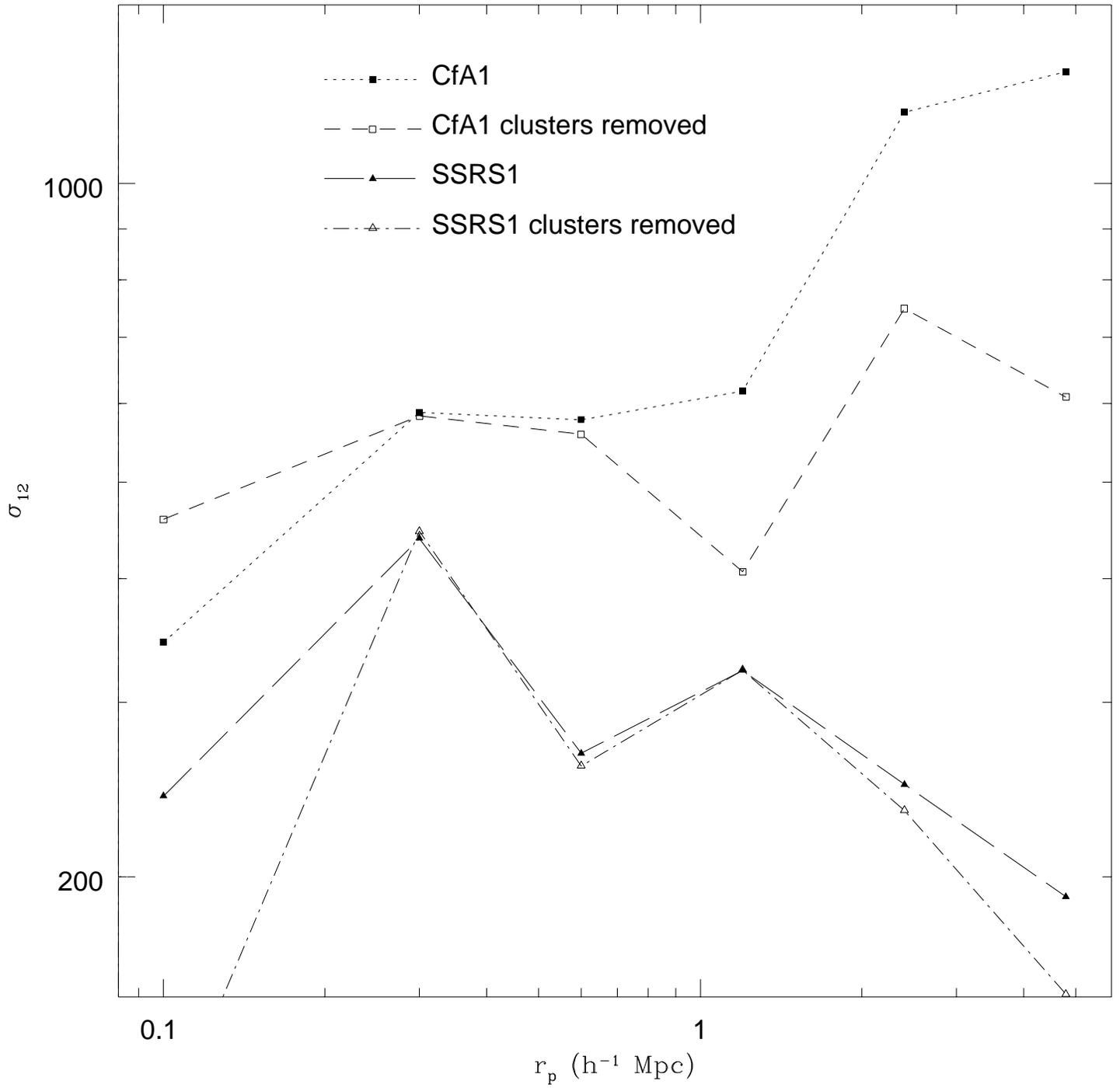
\end{document}